\documentclass[12pt]{iopart}
\usepackage{amsfonts}
\usepackage{amssymb}
\expandafter\let\csname equation*\endcsname=\relax
\expandafter\let\csname endequation*\endcsname=\relax
\usepackage{amsmath}
\usepackage{graphicx}

\begin{document}
\title{Painlev\'e solution of an integral equation}
\author{Y.Y. Atas, E. Bogomolny}
\address{Univ. Paris Sud, CNRS, LPTMS, UMR8626, Orsay F-91405, France}

\begin{abstract}
It is demonstrated that a certain integral equation can be solved using the Painlev\'e equation of third kind. Inversely, a special solution of this Painlev\'e equation can be expressed  as the ratio of two infinite series of spheroidal functions with known coefficients.  
\end{abstract}

\section{Introduction}

We consider an integral equation  in a finite interval for unknown function $g(t)$
\begin{equation}
\int_{-1}^1 K( |x-t|)g(t)\mathrm{d}t=f(x)
\label{equation}
\end{equation} 
where kernel $K(w)$   has the  form 
\begin{equation}
K(w)=w^{\nu}K_{\nu}(\theta w)\ .
\label{operator}
\end{equation}
Here $K_{\nu}(w)$ is the modified Bessel function of the third kind  which obeys the equation
\begin{equation}
K_{\nu}^{\prime \prime }(w)+\frac{1}{w}K_{\nu}^{\prime}(w)-\left (1+\frac{\nu^2}{w^2}\right )K_{\nu}(w)=0
\label{bessel}
\end{equation}
and exponentially decreases at large argument 
\begin{equation}
K_{\nu}(w) \underset{w\to\infty}{\longrightarrow}\sqrt{\frac{\pi}{2 w}}\mathrm{e}^{-w}\ . 
\label{K_asymptotics}
\end{equation}
At small $w$  
\begin{equation}
K_{\nu}(w)\underset{w\to 0}{\longrightarrow} 2^{\nu-1}\Gamma(\nu) w^{-\nu}+2^{-\nu-1} \Gamma(-\nu)w^{\nu}+\Or(w^{2-|\nu|})\ .
\label{small_w} 
\end{equation}
We restrict the discussion to the case $|\nu|<1/2$ though certain results could be generalized to all $\nu>-1/2$.  

Equation \eref{equation} with kernel \eref{operator} appeared in some physical problems. For example, it describes forced convection heat transfer  and contact problems in elasticity (see \cite{belward,mkhitaryan} and references therein). The particular case of \eref{equation}  with  
 $\nu=0$ and imaginary $\theta$ gives the solution of the wave scattering problem with the Dirichlet boundary conditions at the strip $[-1,1]$ (see e.g. \cite{sommerfeld}).  

In \cite{belward} the solution of this equation has been represented as a series of the spheroidal functions  which generalizes the  expansion in terms of Mathieu functions for the case $\nu=0$ obtained in \cite{morse}.   
Equation \eref{equation} belongs to the class of equations whose particular solutions can be obtained by Latta's method \cite{latta} but the resulting equations contain unknown constants.  For $\nu=0$ these constants have been explicitly calculated by Myers \cite{myers}  through a solution of the 
Painlev\'e equation of third kind (a concise summary of Myers' result is given in Appendix B of \cite{tracy}).

The purpose of this note is to show that methods used for $\nu=0$  can be generalized for non-zero $\nu$ ($|\nu|<1/2$) and special  solutions of \eref{equation}  can be obtained from ordinary differential equations (ODE) using the Painlev\'e~III equation more general than the one discussed in \cite{myers}. As a by-product we obtain an explicit formula for a solution of the Painlev\'e~III equation as a ratio of two infinite series of  spheroidal functions with known coefficients. 

The plan of the paper is the following. In Section~\ref{Latta_method}  we use Latta's method to obtain a system of ODE  in $t$ variable for two special solutions of \eref{equation} corresponding to the right-hand side of \eref{equation} equals $\cosh \theta x$ and $\sinh \theta x$. 
These equations contain certain constants which are  determined in Section~\ref{myer_method} by deriving the second system of ODE with respect to variable $\theta$. The condition of the compatibility of these two systems of equations leads to the Painlev\'e~III equation for the constants.  The limiting behaviour of necessary quantities for small and large $\theta$  are derived in Section~\ref{boundary}. Though the discussed methods work only for the above mentioned particular choices of the right-hand side of \eref{equation},  in Section~\ref{embedding_formulae} it is demonstrated that the knowledge of these special solutions permits to investigate  more general problems. In Section~\ref{positivity_relations} we show that kernel \eref{operator} is a positive-definite function which leads to the boundedness of the discussed Painlev\'e III solution at positive arguments.  For completeness, in \ref{series_solution} the series expansion of solutions of integral equation \eref{equation} is present and in \ref{spheroidal_functions} properties of spheroidal functions are briefly described.      
\section{First system of equations}\label{Latta_method}

The method of Latta \cite{latta} can be applied for integral equations of the form \eref{equation} when its kernel, $K(w)$, obeys a differential equation whose coefficients depends linearly on $w$. The kernel $K(w)$ determined in \eref{operator} obeys the equation
\begin{equation}
w K^{\prime \prime}(w)+(1-2\nu)K^{\prime}(w)-w \theta^2 K(w)=0
\label{latta_kernel}
\end{equation}
so the application of Latta's method is straightforward. 

Let us define the integral operator corresponding to \eref{equation}
\begin{equation}
(\Gamma g)(x)\equiv \int_{-1}^1 K( |x-t|)g(t)\mathrm{d}t
\label{Gamma}
\end{equation} 
and its two particular solutions, $g_c(t)$ and $g_s(t)$, with  the right-hand side  equals $\cosh(\theta x)$ and $\sinh(\theta x)$
\begin{equation}
(\Gamma g_c)(x)=\cosh(\theta x)\ ,\quad (\Gamma g_s)(x)=\sinh(\theta x)\ .
\label{latta_definition}
\end{equation}
It is plain that parities of $g_{c,s}(t)$ are fixed: $g_c(-t)=g_c(t)$ and  $g_s(-t)=-g_s(t)$.

An important ingredient of Latta's method is the uniqueness of the solution: when $f(x)=0$ the only solution of \eref{equation} is $g(t)=0$. For kernels like \eref{operator} it is known that if $f(x)$ belongs to $L_2(-1,1)$ then the solution is unique. A simple way to prove it is to use the expansion into a complete set of functions as in \ref{series_solution}.

Changing in \eref{latta_kernel} variable $w$ to $x-t$, multiplying the resulting expression by $g(t)$, and integrating over $t\in[-1,1]$ gives the following identity (the first Latta equation)
\begin{equation}
 (\Gamma \,tg)^{\prime \prime}(x)-\theta^2(\Gamma\, tg)(x)=x\big [ (\Gamma \,g)^{\prime \prime}(x)-\theta^2 (\Gamma\, g)(x)\big ]+(1-2\nu)(\Gamma \,g)^{\prime}(x)=0
\label{Gamma_equation}
\end{equation}
which permits to calculate $(\Gamma \, t g)$ when function $f(x)\equiv (\Gamma\,  g)(x)$ is known and obeys the equation $f^{\prime \prime}-\theta^2 f=0$. 

The second Latta equation is obtained by remarking that  the kernel in \eref{equation} depends only on the difference of arguments. Therefore,  it is easy to check that for arbitrary function $y(t)$ such that $y(\pm 1)=0$
\begin{equation}
(\Gamma\, y)^{\prime}(x)=(\Gamma\, y^{\prime})(x)\ .
\label{latta_deriv}
\end{equation}
To apply this equation it is necessary first to determine the behaviour of function $g(t)$ near ends of the strip,  $t=\pm 1$. Collecting the most singular terms as it is done in Section~\ref{boundary} or using the representation \eref{series_g} gives
\begin{equation}
g_{c,s}\underset{t\to 1}{\longrightarrow} \frac{k_{c,s}}{(1-t)^{\nu+1/2}}\ .
\label{singularities}
\end{equation}
Therefore when $\nu<1/2$  functions 
\begin{equation}
y_{c,s}(t)=(1-t^2)g_{c,s}(t)
\end{equation}
are zero at the both ends of the strip, $y_{c,s}(\pm 1)=0$ and \eref{latta_deriv} is valid for them. 

Equations \eref{Gamma_equation} and \eref{latta_deriv} permit to derive ODE for special solutions \eref{latta_definition} as follows.  
By definition
\begin{equation}
(\Gamma\,g_c)(x)=\cosh \theta x\ . 
\end{equation}
From \eref{Gamma_equation} it follows that $y=(\Gamma\, t g_c)$ obeys the equation
\begin{equation}
y^{\prime\prime }- \theta^2 y=(1-2\nu)\theta\sinh \theta x
\end{equation}
whose odd solution is 
\begin{equation}
y\equiv (\Gamma\, t g_c)(x)=\frac{1-2\nu}{2} x\cosh \theta x +A\sinh \theta x
\end{equation} 
where $A$ is a constant independent on $x$. 

In exactly the same manner the knowledge of $\Gamma\, t g_c$ permits to calculate $\Gamma\, t^2 g_c$
\begin{equation}
(\Gamma\, t^2 g_c)(x)=\frac{(1-2\nu)(3-2\nu)}{8} x^2 \cosh  \theta x + x \sinh \theta x
\Big (\frac{1-2\nu}{2}A-\frac{1-4\nu^2}{8\theta}\Big )
+C\cosh \theta x 
\end{equation}
where $C$ is another constant. 
  
Similarly, with certain constants $B$ and $D$
\begin{equation}
(\Gamma\,g_s)(x)=\sinh \theta x,\qquad 
(\Gamma\, t g_s)(x)=\frac{1-2\nu}{2} x\sinh \theta x +B\cosh  \theta x\ ,
\end{equation}
and 
\begin{equation}
(\Gamma\, t^2 g_s)(x)=\frac{(1-2\nu)(3-2\nu)}{8} x^2 \sinh \theta x + x \cosh \theta x
\Big ( \frac{1-2\nu}{2}B-\frac{1-4\nu^2}{8\theta}\Big )+ D\sinh \theta x\ . 
\end{equation}
Using \eref{latta_deriv} one gets 
\begin{equation}
(\Gamma\, [(1-t^2)g_c]^{\prime})(x)=(\Gamma\, [(1-t^2)g_c])^{\prime}(x)\ .
\end{equation}
From the above relations it follows that
\begin{eqnarray}
 & &(\Gamma\, [(1-t^2)g_c]^{\prime})(x)=-\frac{(1-2\nu)(3-2\nu)}{8}\big [\theta x^2 \sinh \theta x+2x \cosh \theta x\big ]\nonumber\\
&-&\Big (\frac{1-2\nu}{2}A-\frac{1-4\nu^2}{8\theta}\Big )
\big [\theta x\cosh  \theta x+\sinh \theta x\big ]
+(1-C)\theta \sinh \theta  x\ .
\end{eqnarray} 
Expressing the right-hand side through $(\Gamma\, t^2 g_s)$, $(\Gamma t g_c)$ and $(\Gamma g_s)$ one obtains
\begin{equation}
(\Gamma\, [(1-t^2)g_c]^{\prime})(x)=-\theta (\Gamma\, t^2 g_s)(x)+\Big (\theta (B-A)-\frac{3-2\nu}{2}\Big )\, (\Gamma t g_c)(x)
+\rho_1 \,(\Gamma g_s)(x) 
\end{equation} 
where $\rho_1$ is a certain combination of constants $A$, $B$, and $C$. 

As the solution of \eref{equation} is unique, the last equation implies that $g_c$ and $g_s$ have to obey the following ODE
\begin{equation}
(1-t^2)g_c^{\prime}(t)= (\nu+\tfrac{1}{2}-\rho  )tg_c(t)+(\rho_1 -\theta t^2)g_s(t)
\end{equation}
with $\rho=(A-B)\theta+2$.

Repeating these calculations for $g_s$ gives another equation
 \begin{equation}
(1-t^2)g_s^{\prime}(t)= ( \nu+\tfrac{1}{2}+\rho  )tg_s(t)+(\rho_2 -\theta t^2)g_c(t)\ .
\end{equation}
Imposing the condition that near the end $t=1$ functions $g_c(t)$ and $g_s(t)$ have the prescribed singularities (cf. \eref{singularities}) 
\begin{equation}
g_c(t)\underset{t\to 1}{\longrightarrow} \frac{k_c}{(1-t)^{\nu+1/2}},\qquad 
g_s(t)\underset{t\to 1}{\longrightarrow} \frac{k_c\, \eta }{(1-t)^{\nu+1/2}}
\label{limit_g}
\end{equation}   
where $\eta$ is determined from the limit
\begin{equation}
\eta\equiv \frac{k_s}{k_c}=\lim_{t\to 1} \frac{g_s(t)}{g_c(t)}
\label{eta}
\end{equation}
one fixes constants $\rho_1$ and $\rho_2$. 

Finally, we conclude that for $|\nu|<1/2$  functions $g_c(t)$ and  $g_s(t)$ obey the system of ODE
\begin{equation}
\frac{\partial }{\partial t}\left (\begin{array}{c} g_c \\ g_s  \end{array}\right )=
 M\left (\begin{array}{c} g_c \\ g_s \end{array}\right ) 
 \label{first_system}
\end{equation}
with the following $2\times 2$ matrix $M$
\begin{equation}
M=\left (\begin{array}{c c }0 & \theta\\ \theta & 0\end{array}\right ) +\frac{1}{1-t^2}
 \left (\begin{array}{l r } t(\nu+1/2-\rho) & \quad \dfrac{1}{\eta}(\nu+1/2+\rho)\\  \eta (\nu+1/2-\rho)&\quad  t(\nu+1/2+\rho) \end{array}\right ).
 \label{latta_system}
\end{equation}
When $\nu=0$ these equations coincide with the ones in \cite{latta}. 

For the further  use we rewrite the system \eref{first_system} in the form
\begin{eqnarray}
\frac{\partial }{\partial t} \Big [ g_c-\frac{t}{\eta}  g_s\Big ]&=& -\frac{\theta t}{\eta} \, g_c+\Big (\frac{\nu-\frac{1}{2}+\rho}{\eta} + \theta\Big )\, g_s\ ,\nonumber\\
\frac{\partial }{\partial t} \Big [ g_s-t \eta \,g_c\Big ]&=& -\theta t \eta\, g_s+\Big (\eta(\nu-\frac{1}{2}-\rho) + \theta\Big )\,g_c \ .
\label{first_different}
\end{eqnarray}
The terms in the square brackets are zero when $t=\pm 1$ and this system can be obtained directly by using \eref{latta_deriv}.

\section{Second system of equations}\label{myer_method}

To find  constants $\rho$ and $\eta$ we use scaling arguments similar but different to the ones discussed in \cite{myers}. 

Functions $g_c$ and $g_s$ in \eref{latta_definition} depend on $t$ and $\theta$, $g_{c,s}=g_{c,s}(t,\theta)$. We are interesting in finding equations governing the evolution of these function with changing $\theta$.    

Consider instead of system \eref{latta_definition}  a more general system of equations
\begin{eqnarray}
\Gamma G_c \equiv \int_{L_1}^{L_2}|x-t|^{\nu} K_{\nu}( k|x-t|)G_c(t)\mathrm{d}t&=&\cosh \Big [k\big (x-\tfrac{1}{2}(L_1+L_2)\big )\Big ]\ ,\nonumber \\
\Gamma G_s\equiv \int_{L_1}^{L_2} |x-t|^{\nu}K_{\nu}( k|x-t|)G_s(t)\mathrm{d}t&=&\sinh \Big [k \big (x-\tfrac{1}{2}(L_1+L_2)\big )\Big ]\ .
\label{scaling_definition}
\end{eqnarray}
Changing the variable
\begin{equation}
t= \tfrac{1}{2}(L_1+L_2)+\tfrac{1}{2}(L_2-L_1)y
\end{equation}
together with the similar change of $x$    
it is straighforward to check  that functions $G_c(y)$ and $G_s(y)$ are expressed through $g_c(t)$ and $g_s(t)$ defined in \eref{latta_definition} as follows
\begin{equation}
G_{c,s}(y)=\Big (\frac{2}{L_2-L_1}\Big )^{\nu+1} g_{c,s}(z, \theta) 
\label{dependance}
\end{equation}  
where 
\begin{equation}
z=\frac{2}{L_2-L_1}y-\frac{L_2+L_1}{L_2-L_1}\ ,\qquad \theta=\tfrac{1}{2}k(L_2-L_1)\ .
\end{equation}
Consider two functions 
\begin{equation}
\Psi_1(y)=G_s(y)+\eta G_c(y)\ , \qquad \Psi_2(y)=G_s(y)-\eta G_c(y)
\end{equation}
 where $\eta$ is the same as in \eref{eta}. By construction $\Psi_1(L_1)=0$ and  $\Psi_2(L_2)=0$. 
 
From \eref{scaling_definition} it follows that
\begin{equation}
  \Gamma \Psi_{1,2}=\Gamma  G_s\pm \eta \Gamma  G_c=
 \sinh \big (k(x-\tfrac{1}{2}(L_1+L_2)\big ) \pm \eta  \cosh \big (k(x-\tfrac{1}{2}(L_1+L_2)\big )\ . 
 \end{equation} 
As  $\Psi_i(L_i)=0$, the differentiation over $L_i$ of these equations gives
\begin{equation}
\frac{\partial }{\partial L_i}( \Gamma \Psi_i)=\Gamma  \Big(\frac{\partial }{\partial L_i} \Psi_i\Big)\ ,\qquad 
i\textrm{=1,2}\ .
\end{equation}
Using the uniqueness of the solutions,  after simple algebra  we obtain 
\begin{eqnarray}
 \frac{\partial G_s}{\partial L_1}+\eta \frac{\partial G_c}{\partial L_1}& =&-\frac{\theta}{L_2-L_1}(G_c+\eta G_s)\ ,\nonumber\\
 \frac{\partial G_s}{\partial L_2}-\eta \frac{\partial G_c}{\partial L_2}& =&-\frac{\theta}{L_2-L_1}(G_c-\eta G_s)\ . 
 \label{second}
 \end{eqnarray}
Performing the calculations one finds 
\begin{equation}
 \frac{\partial G_{c,s}}{\partial L_i}=\frac{(-1)^{i+1}}{L_2-L_1}\Big (\frac{2}{L_2-L_1}\Big )^{\nu+1} \left((1+\nu)g_{c,s}
 +(z+(-1)^{i}) \frac{\partial g_{c,s}}{\partial z}-\theta \frac{\partial g_{c,s}}{\partial \theta}\right)\ ,
\end{equation} 
where $i=1$ for $G_c$ and $i=2$ for $G_s$. 

Combining all terms together one obtains that \eref{second} take the the form 
\begin{eqnarray}
\theta \eta \frac{\partial g_c}{\partial \theta}&=&-\Big ( \frac{\partial g_s}{\partial z}-\eta z \frac{\partial g_c}{\partial z}\Big )+(\theta +\eta(1+\nu))g_c\ ,\nonumber\\
\theta  \frac{\partial g_s}{\partial \theta}&=&-\Big (z \frac{\partial g_s}{\partial z}-\eta \frac{\partial g_c}{\partial z}\Big )+(\theta \eta+(1+\nu))g_s\ .
\end{eqnarray}
Using \eref{first_different} we conclude that functions $g_{c,s}(t,\theta)$ obey the following system of equations
\begin{equation}
\frac{\partial }{\partial \theta}\left (\begin{array}{c} g_c\\  g_s \end{array}\right )=
 N\left (\begin{array}{c} g_c \\ g_s \end{array}\right ) 
 \label{second_system}
\end{equation}
where matrix $N$ has the form 
\begin{equation}
N=\left ( \begin{array}{cc }\dfrac{1/2+\rho}{\theta} & t\\t & \dfrac{1/2-\rho}{\theta} \end{array}\right )\ .
\label{matrix_N}
\end{equation}
The condition of compatibility of systems of equations \eref{latta_system} and \eref{second_system} gives the equations of zero curvature for matrices $N$ and $M$  
\begin{equation}
\frac{\partial}{\partial t} N-\frac{\partial}{\partial \theta} M=MN-NM\, .
\end{equation}
Direct calculations prove that this equation will be valid provided  $\eta=\eta(\theta)$ and $\rho=\rho(\theta)$ fulfil  the  equations
\begin{equation}
\rho=\frac{\theta}{2\eta}\big (1-\eta^{\prime}-\eta^2\big )\ ,
\label{rho_eq}
\end{equation} 
and 
\begin{equation}
\rho^{\prime}=\frac{\nu+1/2+\rho}{\eta}-(\nu+1/2-\rho)\eta \ .
\end{equation}
Substituting here the previous equation one finds that $\eta$ has to obey equation
\begin{equation}
\frac{\mathrm{d}^2\eta }{\mathrm{d} \theta^2}=\eta^{-1}\Big (\frac{\mathrm{d} \eta }{\mathrm{d} \theta}\Big )^2
-\theta^{-1}\frac{\mathrm{d} \eta }{\mathrm{d} \theta}-\frac{2\nu(1-\eta^2)}{\theta}+\eta^3-\frac{1}{\eta}\ .
\label{painleve}
\end{equation}
This equation is a particular form of  Painlev\'e III equation which has been discussed in \cite{mccoy}. 

From the results of~\ref{series_solution} and \ref{spheroidal_functions} it follows that functions $g_c(\cos \gamma)$ and $g_s(\cos \gamma)$ can be written as the following series of the spheroidal functions
\begin{eqnarray}
g_c(\cos \gamma)&=&\frac{1}{(\sin \gamma)^{2\nu +1}}\sum_{m=0}^{\infty}\mu_{2m}\tilde{X}_{2m}(0)Y_{2m}(\gamma)\ ,\\
g_s(\cos \gamma)&=&\frac{1}{(\sin \gamma)^{2\nu +1}}\sum_{m=0}^{\infty}\mu_{2m+1}\tilde{X}_{2m+1}(0)Y_{2m+1}(\gamma)\ ,
\end{eqnarray} 
where $Y_m(\cos \gamma)$ are defined in \eref{sum_Y}, $X_m(\xi)$ in \eref{third_kind}, and $\mu_n$ in \eref{eigenvalue}.

Function $\eta$ in \eref{eta} which is a solution of the Painlev\'e III equation is given as the ratio of the above series
\begin{equation}
\eta=\dfrac{\sum_{m=0}^{\infty}\mu_{2m+1}\tilde{X}_{2m+1}(0)Y_{2m+1}(0)}{\sum_{m=0}^{\infty}\mu_{2m}\tilde{X}_{2m}(0)Y_{2m}(0)}\ .
\label{series_eta}
\end{equation}
\section{Limiting behaviour}\label{boundary}

In order to use the equations derived in the previous Sections it is necessary to know the behaviour of $g_c(t)$ and $g_s(t)$ defined in \eref{latta_definition} in the limit of small $\theta$ and (or) large $\theta$.

From  \eref{small_w} it follows that at small $\theta$ 
\begin{equation}
K(w)\underset{\theta\to 0}{\longrightarrow} \theta^{-\nu}\frac{\Gamma(\nu)}{2^{1-\nu}}+w^{2\nu}\theta^{\nu} \frac{\Gamma(-\nu)}{2^{1+\nu}}\ .
\label{limit}
\end{equation}
Therefore, to find the necessary  solutions of \eref{latta_definition} when $\theta\to 0$ it is necessary first to solve the equation
\begin{equation}
\int_{-1}^{1}|x-t|^{2\nu}g(t)\mathrm{d}t=f(x)
\end{equation}
when function $f(x)$ equals $1$ or $x$. 

Though the general solution of such equation is known (see  e.g. \cite{williams_2}), solutions with $f(x)=1$ and $f(x)=x$ can easily be obtained by the direct application of Latta's method \cite{latta}. After a simple algebra one finds that the solutions of equations  
\begin{equation}
\int_{-1}^{1}|x-t|^{2\nu}g_0(t)\mathrm{d}t=1,\qquad \int_{-1}^{1}|x-t|^{2\nu}g_1(t)\mathrm{d}t=x
\end{equation}
are 
\begin{equation}
g_0(t)=\frac{\cos(\pi \nu)}{\pi (1-t^2)^{\nu+1/2}}\ , \qquad
g_1(t)=-\frac{\cos(\pi \nu)t}{2\pi \nu (1-t^2)^{\nu+1/2}}\ .
\end{equation}
From \eref{limit} we obtain that
\begin{equation}
g_c(t)\underset{\theta\to 0}{\longrightarrow} \mu_c g_0(t)\ ,\qquad
g_s(t)\underset{\theta\to 0}{\longrightarrow} \mu_s g_1(t)
\end{equation}
where constants $\mu_c$ and $\mu_s$ are determined from the conditions
\begin{equation}
\mu_c\Big ( \theta^{\nu}\frac{\Gamma(-\nu)}{2^{1+\nu}}+
\theta^{-\nu}\frac{\cos(\pi \nu) \Gamma(1/2-\nu) \Gamma(\nu)}{\sqrt{\pi} 2^{1-\nu} \Gamma(1-\nu)} \Big )=1\ ,
\qquad \mu_s \theta^{\nu}\frac{\Gamma(-\nu)}{2^{1+\nu}}=\theta\ . 
\end{equation}
From these relations it follows that 
\begin{equation}
\eta (\theta)\underset{\theta\to 0}{\longrightarrow}-\frac{\mu_s}{2\nu \mu_c}= -\theta^{1-2\nu}\frac{\cos(\pi \nu) \Gamma(1/2-\nu) \Gamma(\nu)}{\sqrt{\pi} 2^{1-2\nu}\nu  \Gamma(1-\nu)\Gamma(-\nu)} -\frac{\theta}{2\nu}\ .
\end{equation}
Using standard formulae for the $\Gamma$-function
\begin{equation}
\Gamma(2\nu)=\frac{2^{2\nu-1}}{\sqrt{\pi}}\Gamma(\nu)\Gamma(\nu+1/2)\ ,\qquad \Gamma(1/2+\nu)\Gamma(1/2-\nu)=\frac{\pi}{\cos(\pi\nu)}\ ,
\end{equation}
one gets that the limiting behaviour of $\eta(\theta)$ at small $\theta$ is
\begin{equation}
\eta (\theta) \underset{\theta\to 0}{\longrightarrow} B(2\theta)^{1-2\nu}-\frac{\theta}{2\nu}=
\left \{ \begin{array}{cc} B(2\theta)^{1-2\nu},& 0<\nu<1/2\\-\theta/(2\nu), & -1/2<\nu<0 \end{array}\right .  
\label{asymptotics_eta}
\end{equation} 
where
\begin{equation}
B=2^{-3(1-2\nu)}\frac{\Gamma^2(\nu)}{\Gamma^2(1-\nu)\Gamma(2\nu)}\ .
\label{our_B}
\end{equation}
To find  the behaviour of solutions when $\theta\to\infty$ it is convenient to consider instead of functions  $g_{c,s}$  \eref{latta_definition} a solution $g_{-}(v)$ corresponding to the integral equation 
\begin{equation}
\int_{-1}^{1}|u-v|^{\nu}K_{\nu}(\theta |u-v|)g_{-}(v)\mathrm{d}v=\mathrm{e}^{-\theta u}\ . 
\label{wiener_hopf_minus}
\end{equation} 
To find  asymptotic behaviour of $g_{-}$ for large $\theta$ we first consider  the  equation  
\begin{equation}
\int_0^{\infty}|u-v|^{\nu}K_{\nu}(|u-v|)g_{0}(v)\mathrm{d}v=\mathrm{e}^{-u}\ .
\label{wiener_hopf}
\end{equation}
Its solution can be calculated either by the Wiener-Hopf method \cite{noble} using known Fourier transform of the kernel
\begin{equation}
\int_{-\infty}^{\infty}\mathrm{e}^{\mathrm{i}px}
K_{\nu}(\theta x)x^{\nu}\mathrm{d}x=\frac{(2\theta)^{\nu}\sqrt{\pi}\Gamma(\nu+1/2)}{(p^2+\theta^2)^{\nu+1/2}}
\label{fourier_kernel}
\end{equation} 
or by expansion formulae  analogous to the ones described in \ref{series_solution} but for half-line integration \cite{belward_2} 
\begin{eqnarray}
&& n!\int_0^{\infty} |x-t|^{\nu} K_{\nu}\big (\tfrac{1}{2}|x-t|\big ) t^{-\nu-1/2} \mathrm{e}^{-t/2} L_n^{-\nu-1/2}(t)\mathrm{d}t\nonumber\\
&&= 
\sqrt{\pi} \Gamma(\nu+1/2) \Gamma(n+1/2-\nu) \mathrm{e}^{-x/2} L_n^{-\nu-1/2}(x) 
\end{eqnarray}
where $L_n^{\lambda}(x)$ with $n=0,1,\ldots$ are the Laguerre polynomials. 

One gets
\begin{equation}
g_0(v)=C_{\nu}v^{-\nu-1/2}\mathrm{e}^{-v}\ ,\qquad C_{\nu}=\frac{\sqrt{2}}{\pi^{3/2}}\cos \pi \nu\ .
\label{W_H_solution}
\end{equation}
Using this solution it is easy to show that the dominant approximation for the solution of \eref{wiener_hopf_minus} when $\theta\to \infty$ has the form 
\begin{equation}
g_{-}^{(0)}(t)=C_{\nu}\sqrt{\theta}(1+t)^{-\nu-1/2}\mathrm{e}^{-t\theta}\ . 
\label{approximation}
\end{equation}
Straightforward transformations give
\begin{equation}
\int_{-1}^1 |x-t|^{\nu}K_{\nu}(\theta |x-t|)g_{-}^{(0)}(t)\mathrm{d}t= F_1(x)-F_2(x)
\end{equation}
where
\begin{equation}
F_1(x)=\int_{-1}^{\infty} |x-t|^{\nu}K_{\nu}(\theta |x-t|)g_{-}^{(0)}(t)\mathrm{d}t\ ,\; 
F_2(x)=\int_{1}^{\infty} |x-t|^{\nu}K_{\nu}(\theta |x-t|)g_{-}^{(0)}(t)\mathrm{d}t\ .
\end{equation}
Changing the variable in the first integral as $t=-1+v/\theta$ and using  \eref{wiener_hopf} leads to 
\begin{equation}
F_1(x)=\mathrm{e}^{-\theta x}\ .
\end{equation}
In the integral for $F_2(x)$ it is convenient to put $t=1+s/\theta$. It this way one obtains
\begin{equation}
F_2(x)= \theta^{-\nu-1/2} C_{\nu}\int_{0}^{\infty}\Big (2+\frac{s}{\theta}\Big )^{-\nu-1/2}
(s+\theta(1-x))^{\nu}K_{\nu}(s+\theta(1-x)) \mathrm{e}^{-\theta-s}\mathrm{d}s\ .
\end{equation}
When $\theta$ is large one can drop the term $s/\theta$ in the first bracket and use the asymptotic form of the $K_{\nu}$ function \eref{K_asymptotics}
\begin{equation}
F_2(x)\underset{\theta\to\infty}{\longrightarrow} 
\theta^{-\nu-1/2} 2^{-\nu-1}\sqrt{\pi}\mathrm{e}^{-2\theta+\theta x}  C_{\nu}\int_{0}^{\infty}(s+\theta(1-x))^{\nu -1/2} \mathrm{e}^{-2s}\mathrm{d}s\ .
\end{equation}
This formula means that close to right-hand side, $x=1$, the additional contribution related with the integration in finite limits is  
\begin{equation}
F_2(x)\approx \theta^{-\nu-1/2} 2^{-\nu-1}\sqrt{\pi}\mathrm{e}^{-2\theta+\theta x}  C_{\nu}\int_{0}^{\infty}s^{\nu -1/2} \mathrm{e}^{-2s}\mathrm{d}s=\mathrm{e}^{\theta x}\delta 
\end{equation}
where
\begin{equation}
\delta=\frac{1}{2\pi}\theta^{-\nu-1/2} 2^{-2\nu}\cos(\pi \nu) \mathrm{e}^{-2\theta}\Gamma(\nu+1/2)\ . 
\end{equation}
Using \eref{wiener_hopf_minus} it is obvious that to cancel contribution $F_2(x)$ it is necessary to modify the approximation \eref{approximation} as follows
\begin{equation}
g_{-}(t)\underset{\theta\to \infty}{\longrightarrow} C_{\nu}\sqrt{\theta}\Big [(1+t)^{-\nu-1/2}\mathrm{e}^{-t\theta}+
\delta (1-t)^{-\nu-1/2}\mathrm{e}^{t\theta}\Big ]\ . 
\label{main_approximation}
\end{equation}
This function consists of two terms. The first dominates close to $t=-1$ and the second near $t=1$. According to \eref{eta} function $\eta(\theta)$ is determined from the limit
\begin{equation}
\eta(\theta)=\lim_{t\to 1}\frac{g_{-}(-t)-g_{-}(t)}{g_{-}(-t)+g_{-}(t)}\ .
\end{equation} 
Using \eref{main_approximation} we find that 
\begin{equation}
\eta(\theta)\underset{\theta\to \infty}{\longrightarrow} 1-\frac{\cos(\pi\nu) }{\pi} \Gamma(\nu+1/2)2^{-2\nu}\theta^{-\nu-1/2}\mathrm{e}^{-2\theta}\ . 
\label{large_theta_eta}
\end{equation} 
\section{Embedding formulae}\label{embedding_formulae}

In the previous Sections the integral equation \eref{equation}  has been transformed into two systems of ODE \eref{first_system}, \eref{latta_system} and \eref{second_system}, \eref{matrix_N}  but only for very special right-hand sides of this equation, $f(x)=\cosh (\theta x)$ and $f(x)=\sinh (\theta x)$  (cf. \eref{latta_definition}).  Nevertheless, the knowledge of these two special solutions permits to find solutions of \eref{equation} for more general cases \cite{tracy}, \cite{williams}. 

Let us consider the equation of the form 
\begin{equation}
(\Gamma g)(x)=\mathrm{e}^{-\theta z x}
\label{general}
\end{equation} 
with the same kernel as in \eref{operator}. To find its solution, $g(t)$, it is convenient  to define symmetric and antisymmetric combinations of $g_{c,s}$
\begin{equation}
g_{\pm}(t)=g_c(t)\pm g_s(t)\ .
\end{equation} 
We look for solution of \eref{general} in the form 
\begin{equation}
\Psi(t)=g(t)+a_{+} g_{+}(t)+a_{-} g_{-}(t)
\end{equation}
where  $a_{+}$ and $a_{-}$ are two constants independent on $t$ which have to be determined from two conditions
\begin{equation}
\lim_{t\to\pm 1}\Psi(t)=0\ . 
\label{limit_Psi}
\end{equation} 
The explicit form of these constants will be presented later (see \eref{a_pm}). 

Assuming that \eref{limit_Psi} is fulfilled, one concludes as in Section~\ref{Latta_method} that
\begin{equation}
(\Gamma  \Psi(x))^{\prime}=(\Gamma \Psi^{\prime})(x)\ .
\end{equation} 
The right-hand side of this equation is calculated from the definition of $\Psi$ and
\begin{equation}
(\Gamma  \Psi(x))^{\prime}=-\theta z\mathrm{e}^{-\theta z x} +\theta  a_{+} \mathrm{e}^{\theta x} -\theta  a_{-} \mathrm{e}^{-\theta x}
=(\Gamma [  -z \theta g +a_{+}\theta g_{+} -a_{-}\theta g_{-}])(x)\ .
\end{equation} 
Due to the uniqueness of the solution,  $\Psi(t)$ obeys the equation
\begin{equation}
\Psi^{\prime}(t)=-z\theta \Big (\Psi(t)-a_{+}g_{+}(t)- a_{-}g_{-}(t)\Big )+a_{+} \theta g_{+}(t)+ a_{-} \theta g_{-}(t)
\end{equation}
which is equivalent to 
\begin{equation}
\theta^{-1}\Psi^{\prime}(t)+z \Psi(t)=(z +1)a_{+} g_{+}(t) +(z  -1 )a_{-}  g_{-}(t)
\label{Psi_equation}
\end{equation}
whose solution is straightforward if constants $a_{\pm}$ are known. 

To find them one can proceed as follows \cite{williams}. Let us calculate the following quantities 
\begin{equation}
(g_{\pm }\Gamma \Psi)\equiv \int_{-1}^1 \int_{-1}^1g_{\pm}(x)K(x-t)\Psi(t)\mathrm{d}t\mathrm{d}x
\end{equation}
by two different methods, the first by applying the operator $\Gamma$ on $g_{\pm}$ and the second by applying it on $\Psi$. In such a manner one gets 
\begin{equation}
\int_{-1}^1 \Psi(t)\mathrm{e}^{\pm \theta t}\mathrm{d}t = \int_{-1}^{1}g_{\pm}(t)\Big ( \mathrm{e}^{-z \theta t}+
a_{+} \mathrm{e}^{\theta t}+a_{-} \mathrm{e}^{-\theta t}\Big )\mathrm{d}t\ .
\end{equation}
Introduce  the Laplace transform of functions $g_{c,s}$
\begin{equation}
\hat{G}_c(p)=\int_{-1}^1g_{c}(t)\mathrm{e}^{p\theta t}\mathrm{d}t\ , \qquad
\hat{G}_s(p)=\int_{-1}^1g_{s}(t)\mathrm{e}^{p\theta t}\mathrm{d}t\ .
\label{laplace}
\end{equation}
Due to the symmetry properties of $g_{c,s}$, $\hat{G}_c(-p)=\hat{G}_c(p)$ and  $\hat{G}_s(-p)=-\hat{G}_s(p)$. 

Using \eref{laplace} we obtain
\begin{eqnarray}
\int_{-1}^1 \Psi(t)\mathrm{e}^{\theta t}\mathrm{d}t&=&G(-z)+a_{+} G(1)+a_{-} G(-1)\ ,\\
\int_{-1}^1 \Psi(t)\mathrm{e}^{-\theta t}\mathrm{d}t&=&G(z)+a_{+} G(-1)+a_{-} G(1)
\end{eqnarray}
where  $G(p)=\hat{G}_c(p)+\hat{G}_s(p)$. 
 
On the other hand, multiplying \eref{Psi_equation} by $\mathrm{e}^{\pm \theta t}$,  integrating the results from $-1$ to $1$, and taking into account that $\Psi(\pm 1)=0$ one finds
\begin{eqnarray}
(z-1)\int_{-1}^1  \mathrm{e}^{\theta t}\Psi(t)\mathrm{d}t&=&(z+1) a_{+}G(1)+(z-1)a_{-}G(-1)\ ,\\
(z+1)\int_{-1}^1  \mathrm{e}^{-\theta t}\Psi(t)\mathrm{d}t&=&(z+1) a_{+}G(-1)+(z-1)a_{-}G(1)\ .
\end{eqnarray}
These  and  previous equations permits to calculate constants $a_{\pm}$
\begin{equation}
a_{\pm}=-\frac{(1\mp z)G(\mp z)}{2G(1)}\ .
\label{a_pm}
\end{equation}
Combining the above  formulae one finds that \eref{Psi_equation} takes the form
\begin{equation}
\theta^{-1}\Psi^{\prime}(t)+z \Psi(t)=\frac{z^2-1}{G(1)}\big [ \hat{G}_c(z) g_{s}(t) -\hat{G}_s(z)  g_{c}(t)\big ]
\label{equation_Psi}
\end{equation}  
whose solution obeying $\Psi(\pm 1)=0$ is
\begin{equation}
\Psi(t)=\frac{(z^2-1)\theta}{G(1)} \mathrm{e}^{-z\theta t}\Big [\hat{G}_c(z) \int_{-1}^t g_{s}(y)\mathrm{e}^{z\theta y}\mathrm{d}y -\hat{G}_s(z)\int_{-1}^t g_{s}(y)\mathrm{e}^{z\theta y}\mathrm{d}y\Big ]\ .
\label{Psi_Z}
\end{equation}
The above formulae relate the general plane-wave  solution \eref{general} to two special cases corresponding to $z=\pm 1$. Such formulae are called embedding formulae in the theory of diffraction and can be derived in more general settings (see e.g. \cite{shanin} and references therein). 
\section{Positivity relations}\label{positivity_relations}

Equation \eref{Psi_Z} represents the solution of \eref{general}. As for usual diffraction problems it is physically clear (and can be proved e.g. by using the series expansions as in \ref{series_solution}) that such solution  exist   for all $\theta>0$ and any $z$ but \eref{Psi_Z} contains $G(1)$ in the numerator and in order to the solution remains finite it is necessary that this quantity is always non-zero. 

By the construction $G(1)$ is the Laplace transform of the sum $g_c+g_s$ 
\begin{equation}
G(1)=\int_{-1}^1(g_{c}(t)+g_{s}(t))\mathrm{e}^{\theta t}\mathrm{d}t=\int_{-1}^1g_{c}(t)\cosh(\theta t)\mathrm{d}t
+\int_{-1}^1g_{s}(t)\sinh(\theta t)\mathrm{d}t\ .
\end{equation} 
By definition, $g_c(t)$ and $g_s(t)$ are solutions of \eref{latta_definition} thus 
\begin{equation}
\cosh(\theta x)=(\Gamma g_c)(x),\quad \sinh(\theta x)=(\Gamma g_s)(x)
\end{equation}
and 
\begin{equation}
G(1)=\int_{-1}^1\mathrm{d}t \int_{-1}^1 g_c(t)K(|x-t|)g_c(x)\mathrm{d}x+
\int_{-1}^1\mathrm{d}t \int_{-1}^1 g_s(t)K(|x-t|)g_s(x)\mathrm{d}x
\end{equation}
with the kernel $K(w)$ defined in \eref{operator}. An important property of such kernel is that its Fourier transform, $\hat{K}(p)$, presented \eref{fourier_kernel} is strictly positive for all $p$ and $\nu>-1/2$. It means that function $K(w)$ is a positive definite function and the double integral
\begin{equation}
\int_{-1}^1\mathrm{d}t \int_{-1}^1 f(t)K(|x-t|)f(x)\mathrm{d}x  \equiv \frac{1}{2\pi}\int_{-\infty}^{\infty}
\Big | \int_{-1}^1   f(t)\mathrm{e}^{\mathrm{i}pt}\mathrm{d}t \Big |^2 \hat{K}(p)\mathrm{d}p
\end{equation} 
is positive  for any function $f(x)$. Therefore for all $\theta>0$
\begin{equation}
G(1)>0
\label{positivity}
\end{equation}
and the solution \eref{Psi_Z} is finite  for all $z$.

The same positivity condition permits also to establish the positivity and finiteness of $\eta(\theta)$. Formally, this function is defined as the limit when $t\to 1$ of the ratio of two solutions \eref{eta} and it is not evident that it remains finite for all $\theta>0$. 

Let us define two functions           
\begin{equation}
\hat{F}_c(p)=\int_{-1}^1 tg_{c}(t)\mathrm{e}^{p\theta t}\mathrm{d}t\ , \qquad
\hat{F}_s(p)=\int_{-1}^1 tg_{s}(t)\mathrm{e}^{p\theta t}\mathrm{d}t
\label{laplace_t}
\end{equation}
which differ from \eref{laplace} by factor $t$ in the integrand. 

Multiplying \eref{first_different} by  $\mathrm{e}^{z\theta t}$ and integrating the both parts over the interval $[-1,1]$ one finds 
\begin{eqnarray}
\Big [\frac{\nu-1/2-\rho}{\theta}+\frac{1}{\eta} \Big ]\hat{G}_c(z)+\frac{z}{\eta} \hat{G}_{s}(z)&=&z\hat{F}_c(z)+\hat{F}_{s}(z)\ ,\nonumber \\
\Big [\frac{\nu-1/2+\rho}{\theta}+\eta \Big ]\hat{G}_s(z)+z \eta \hat{G}_{c}(z)&=&\hat{F}_c(z)+z\hat{F}_{s}(z)\ .
\end{eqnarray}
From \eref{second_system} and \eref{matrix_N} it follows that
\begin{eqnarray}
\frac{\partial}{\partial \theta}\hat{G}_c(z)-\frac{1/2+\rho}{\theta}\hat{G}_c(z)&=&\hat{F}_s(z)+z\hat{F}_c(z)\ ,\nonumber\\
\frac{\partial}{\partial \theta}\hat{G}_s(z)-\frac{1/2-\rho}{\theta}\hat{G}_s(z)&=&\hat{F}_c(z)+z\hat{F}_s(z)\ .
\end{eqnarray}
Combining these relations one concludes that
\begin{equation}
\frac{\partial}{\partial \theta}\hat{G}_c(z)-\Big [\frac{\nu}{\theta}+\frac{1}{\eta}\Big ]\hat{G}_c(z)=\frac{z}{\eta}\hat{G}_s(z)\ , \qquad
\frac{\partial}{\partial \theta}\hat{G}_s(z)-\Big [\frac{\nu}{\theta}+\eta \Big ]\hat{G}_s(z)= z\eta \hat{G}_c(z)\ .
\end{equation}
Put in the last expressions $z=1$. Then $G(1)=\hat{G}_c(1)+\hat{G}_s(1)$ obeys the equation
\begin{equation}
\frac{\partial}{\partial \theta}G(1)=\Big [\frac{\nu}{\theta}+\eta+\frac{1}{\eta}\Big ]G(1)
\end{equation}
whose solution is
\begin{equation}
G(1)=C^{\prime}(\nu) \, \theta^{\nu}\,\exp \int^{\theta}(\eta(\theta^{\prime})+\eta^{-1}(\theta^{\prime}))\mathrm{d}\theta^{\prime}
\label{G_1}
\end{equation}
where $C^{\prime}(\nu)$ is a constant. 

Using \eref{limit_g},  \eref{second_system}, and \eref{matrix_N} with $t=1$ one concludes that
\begin{equation}
\frac{\partial}{\partial \theta} k_c=\frac{1/2+\rho}{\theta} k_c +k_c\eta. 
\end{equation}
Solving it and using \eref{rho_eq} we finds
\begin{equation}
k_c^2 =C^{\prime \prime}(\nu)\frac{\theta}{\eta}\exp \int^{\theta}(\eta(\theta^{\prime})+\eta^{-1}(\theta^{\prime}))\mathrm{d}\theta^{\prime}
\end{equation}
with another constant $C^{\prime\prime}(\nu)$. 

Comparison of this expression with \eref{G_1} proves that
\begin{equation}
\eta \, k_c^2=C(\nu)\, \theta^{1-\nu}G(1)\ . 
\label{main_positive}
\end{equation}
To find the constant of proportionality, $C(\nu)$,  we compare the both sides of this equation when $\theta\to 0$.

From Section~\ref{boundary} it follows that for $|\nu|<1/2$ and $\theta\to 0$ the dominant contribution to  $G(1)$ is due to $\hat{G}_c(1)$ and it is straightforward to check that 
\begin{equation}
C(\nu)=\frac{\cos \pi \nu}{2^{\nu+1}\pi^{3/2}\Gamma(1/2-\nu)}\ .
\label{constant_C}
\end{equation}
For $|\nu|<1/2$, $C(\nu)>0$ and due to the positivity of $G(1)$  it follows from \eref{main_positive}  that $\eta>0$ for all positive $\theta>0$. As the only possible moving singularities of $\eta$ are poles this reasoning proves that $\eta$ is bounded for all $\theta>0$ (the behaviour of $\eta$ at large $\theta$ is fixed by \eref{large_theta_eta}).  

The list of solutions of the Painlev\'e III equation \eref{painleve} which remain bounded as $\theta\to \infty$ 
along the real axis has been presented in \cite{mccoy}. Such  solutions form one parameter family parametrized by parameter $\lambda$  which determined the large $\theta$ behaviour of solutions
\begin{equation}
\eta(\theta)\underset{\theta\to \infty}{\longrightarrow}1-\lambda \Gamma(\nu+1/2)2^{-2\nu}\theta^{-\nu-1/2}\mathrm{e}^{-2\theta}\ . 
\label{large_theta}
\end{equation} 
Bounded solutions are characterized by the following behaviour at small $\theta$ 
\begin{equation}
\eta(\theta)\underset{\theta\to 0}{\longrightarrow} (2\theta)^{\sigma}B +(2\theta) B_1+  (2\theta)^{1+2\sigma}B_2 +(2\theta)^{2-\sigma}B_3+\Or(\theta^{2+\sigma})
\end{equation}
where $\sigma$ (restricted by inequalities  $-1<\mathrm{Re}\, \sigma <1$) is related with $\lambda$ as follows 
\begin{equation}
\sigma=\frac{2}{\pi}\arcsin (\pi \lambda)\ .
\label{sigma_lambda}
\end{equation}
and coefficients $B_j$ are 
\begin{equation}
B_1=-\frac{\nu}{(1-\sigma)^2},\quad B_2=B^2\frac{\nu}{(1+\sigma)^2},\quad B_3=\frac{1}{16 B(1-\sigma)^4}(4\nu^2-(1-\sigma)^2)
\label{series_B}
\end{equation}
with $B=B(\sigma,\nu)$ being a function of $\sigma$ and $\nu$
\begin{equation}
B(\sigma,\nu)=2^{-3\sigma}\frac{\Gamma^2((1-\sigma)/2)\Gamma((1+\sigma)/2+\nu)}{\Gamma^2((1+\sigma)/2)\Gamma((1-\sigma)/2+\nu)}\ .
\label{mccoy_B}
\end{equation}
When $0<\nu<1/2$ the solution of the integral equation \eref{equation} discussed above corresponds to a particular case of general theory in \cite{mccoy} with $\sigma=1-2\nu$ (cf. \eref{our_B}, \eref{mccoy_B}, and \eref{large_theta_eta}). It seems that the case when $-1/2<\nu<0$ is missed from the analysis of \cite{mccoy} as  $1<\sigma=1-2\nu<2$. For such choice of $\sigma$ the term $B_3$ in \eref{series_B} is identically zero and the asymptotics \eref{asymptotics_eta} represents the two leading terms when $\theta\to 0$. Though in this case the dominant behaviour at small $\theta$ is linear in $\theta$,   $\eta\to -\theta/(2\nu)$, its asymptotics at large $\theta$ is given by the same formula \eref{large_theta} but with value of $\lambda$ obtained by the inversion of \eref{sigma_lambda} (see \eref{large_theta_eta})  
\begin{equation}
\lambda=\frac{\cos(\pi \nu)}{\pi}\ .
\end{equation} 
\section{Conclusion}

We demonstrate that special solutions of  the integral equation \eref{equation} can be obtained from ordinary differential equations \eref{first_system}, \eref{latta_system} or \eref{second_system}, \eref{matrix_N} with entering constants calculated  using the Painlev\'e~III equation \eref{painleve} with known asymptotics at small and large arguments. Our finding generalizes the results obtained in \cite{myers} for the special case $\nu=0$.  An interesting consequence of our investigation is the expression \eref{series_eta} for a solution of the Painlev\'e III equation with asymptotics given by \eref{asymptotics_eta} and \eref{large_theta_eta} as the ratio of two infinite series of spheroidal functions with known coefficients.  The positive--definiteness of the kernel forces this  Painlev\'e~III solution to be bounded for all positive arguments thus giving another proof of connection formulae for the Painlev\'e~III equation \cite{mccoy} in a special case.    

\ack

One of the author (EB) is greatly indebted to A. Its for pointing out the important reference \cite{myers}.
\appendix

\section{Series solution of the integral equation}\label{series_solution}

Let  $\mu_m$ and $Y_m(\gamma)$ be generalized eigenvalues and eigenfunctions  of operator \eref{operator}
\begin{equation}
\mu_m \int_{0}^{\pi}(\sin \gamma)^{-2\nu} K( |\cos \beta -\cos \gamma \, |) Y_m(\gamma)\mathrm{d}\gamma=Y_m(\beta)\ .
\label{eigen_equation}
\end{equation} 
As the symmetrized version of the kernel has the form
\begin{equation}
\tilde{K}(\beta, \gamma)=(\sin \beta)^{-\nu}K(|\cos \beta -\cos \gamma \, |)(\sin \gamma)^{-\nu}
\end{equation}
eigenfunctions $Y_m(\gamma)$ form an orthogonal system of functions 
\begin{equation}
\int_{0}^{\pi}(\sin \gamma)^{-2\nu} Y_m(\gamma)Y_n(\gamma)\mathrm{d}\gamma=N_m\delta_{mn}
\label{orthogonality}
\end{equation}
where $N_m$ is the normalization constant.

If a function $f(x)$ is expanded into a series of functions $Y_m(\gamma)$    
\begin{equation}
f(\cos \gamma)=\sum_m a_m Y_m(\gamma)
\label{series_f}
\end{equation}
then the formal solution of the integral  equation \eref{equation} is 
\begin{equation}
g(\cos \gamma )= (\sin \gamma)^{-2\nu -1}\sum_{m=1}^{\infty} \mu_m a_m Y_m(\gamma)\ .
\label{series_g}
\end{equation}
To find explicitly  eigenvalues and eigenfunctions of \eref{eigen_equation} it was noted in  \cite{belward} that the function $K(\sqrt{x^2+y^2})$ plays the role of the Green function of the equation 
\begin{equation}
\left ( \frac{\partial^2}{\partial x^2} + \frac{\partial^2}{\partial y^2}
-\frac{2\nu}{y}\frac{\partial}{\partial y}-\theta^2 \right )\Psi(x,y)=0\ .
\label{partial}
\end{equation}
Therefore the function 
\begin{equation}
\Psi^{(\mathrm{ref})}(x,y)=\int_{-1}^{1}g(t)K(\sqrt{(t-x)^2+y^2})\mathrm{d}t
\label{reflected}
\end{equation}
is a uni-valued  solution of \eref{partial} in all points of the $(x,y)$-plane except the strip $[-1,1]$ which   exponentially  decays at large distances
\begin{equation}
\Psi^{(\mathrm{ref})}(r\cos \phi,\, r\sin \phi)\underset{r\to\infty}{\longrightarrow} 
\sqrt{\frac{2}{\theta \pi}}r^{\nu-1/2}\, 
\mathrm{e}^{-\theta r}\int_{-1}^{1}g(t)\mathrm{e}^{\theta t \cos \phi}\mathrm{d}t \ .
\end{equation} 
Consider a Dirichlet-type problem of finding the solution of \eref{partial} in the form
\begin{equation}
\Psi(x,y)=\Psi^{(\mathrm{inc})}(x,y)+\Psi^{(\mathrm{ref})}(x,y)
\end{equation}
where  $\Psi^{(\mathrm{inc})}(x,y)$ also obeys \eref{partial} and the total field $\Psi(x,0)$ is zero at the interval $[-1,1]$
\begin{equation}
\Psi(x,0)=0\ ,\qquad -1\leq x \leq 1\ .
\label{dirichlet}
\end{equation}
It means that unknown function $g(t)$ obeys \eref{equation}  with $f(x)=-\Psi^{(\mathrm{inc})}(x,0)$. 

Equation \eref{partial} permits the separation of variables in the elliptic coordinates
\begin{equation}
x=\cosh \xi \cos \gamma\ ,\qquad y=\sinh \xi \sin \gamma\ . 
\label{elliptic_coordinates}
\end{equation}
It is plain that
\begin{equation}
 \frac{\partial^2}{\partial x^2} + \frac{\partial^2}{\partial y^2}=\frac{2}{\cosh 2\xi-\cos 2\gamma}\Big ( \frac{\partial^2}{\partial \xi^2} + \frac{\partial^2}{\partial \gamma^2}\Big )\ ,
\end{equation}
and
\begin{equation}
\frac{\partial}{\partial y}=\frac{2 }{\cosh 2\xi-\cos 2\gamma}\Big (\cosh \xi \sin \gamma\frac{\partial}{\partial \xi}+
\sinh \xi \cos \gamma \frac{\partial}{\partial \gamma}\Big )\ .
\end{equation}
If $\Psi(x,y)=X(\xi)Y(\gamma)$ then   $Y(\gamma)$ and $X(\xi)$ obey 
\begin{equation}
Y_m^{\prime  \prime}-2\nu \cot \gamma \, Y_m^{\prime }-(\alpha_m-\tfrac{1}{2}\theta^2 \cos 2\gamma)Y_m=0
\label{spheroidal}
\end{equation} 
and
\begin{equation}
X_m^{\prime  \prime}-2\nu \coth \xi \, X_m^{\prime }+(\alpha_m -\tfrac{1}{2}\theta^2 \cosh 2\xi)X_m=0\ . 
\label{modified}
\end{equation}  
The separation constant, $\alpha_m$, is chosen to  ensure the symmetry properties of solutions
\begin{equation}
 Y_m(-\gamma)=Y_m(\gamma)\ ,\qquad Y_m(\gamma)=Y_m(\gamma+\pi)\ .
\label{symmetry}
\end{equation}
The solution $X_m(\xi)$ is fixed by its normalization at infinity
\begin{equation}
X_m(\xi)\underset{\xi\to\infty}{\longrightarrow} C\mathrm{e}^{\xi (\nu -1/2)}\exp \left (-\frac{\theta}{2}\mathrm{e}^{\xi }\right )
\label{X_asymptotics}
\end{equation}
where $C$ is a constant. 

The above equations are  particular cases of spheroidal equation \cite{meixner}. $Y_m(\gamma)$ is even periodic angular  solution of \eref{spheroidal} and $X_m(\xi)$ is the radial  solution of the modified equation \eref{modified} decreasing at the infinity. It is known that $Y_m$ are orthogonal functions on interval $[0,\ \pi]$ as in \eref{orthogonality} and they form a complete set in the space of even functions on this interval. 

Therefore, the reflected uni-valued field  in the $(x,y)$-plane cut along the interval $[-1,1]$,  decaying at infinity (cf. \eref{X_asymptotics}), and obeying \eref{dirichlet} with $f(x)$ as in \eref{series_f} can be represented as a formal series
\begin{equation}
\Psi^{(\mathrm{ref})}(\xi,\gamma)=\sum_m a_m \frac{X_m(\xi)}{X_m(0)}Y_m(\gamma)\ . 
\end{equation}
As for the usual Dirichlet problem the value of $g(t)$ in \eref{reflected} is related with the normal derivative of  $\Psi^{(\mathrm{ref})}$ at the strip $[-1,1]$. At small $w$  
\begin{equation}
K_{\nu}(w)\underset{w\to 0}{\longrightarrow} w^{-\nu}\frac{\Gamma(\nu)}{2^{1-\nu}}+w^{\nu}\frac{\Gamma(-\nu)}{2^{1+\nu}}\ . 
\end{equation}
As $w=\sqrt{(x-t)^2+y^2}$ and for $\nu <1/2$
\begin{equation}
\lim_{y\to 0+} \frac{y^{1-2\nu}}{(x^2+y^2)^{1-\nu}}= \frac{\sqrt{\pi}\Gamma(1/2-\nu)}{\Gamma(1-\nu)}\delta(x)
\end{equation}
one gets that \cite{belward}
\begin{equation}
g(x)=-\frac{2^{\nu}\theta^{-\nu }}{\sqrt{\pi} \Gamma(1/2-\nu)}\lim_{y\to 0} y^{-2\nu}
\frac{\partial }{\partial y}\Psi^{(\mathrm{ref})}(x,y)\ .
\end{equation}
The behaviour of the decaying solution $X_m(\xi)$ at small $\xi$ follows from \eref{modified}
\begin{equation}
X_m(\xi)\underset{\xi\to 0}{\longrightarrow} A_m+B_m \xi^{1+2\nu}\ .
\label{small_X}
\end{equation}  
For $x$ in the strip $[-1,1]$ and small $y$, $y=\xi \sin \gamma$. Therefore  
\begin{equation}
g(\cos \gamma)=-\frac{2^{\nu}(2\nu+1)\theta^{-\nu}}{\sqrt{\pi} \Gamma(1/2-\nu)} (\sin \gamma)^{-2\nu-1}\sum_m a_m 
\frac{B_m}{A_m}Y_m(\gamma)\ .
\end{equation} 
It means that eigenfunctions of operator \eref{eigen_equation} equal $Y_m$ and the corresponding eigenvalues are 
\begin{equation}
\mu_m=-\frac{2^{\nu}\theta^{-\nu }(2\nu+1)B_m}{\sqrt{\pi} \Gamma(1/2-\nu)A_m}
\label{eigenvalue}
\end{equation}
where $A_m$ and $B_m$ are determined from the dominant behaviour at small $\xi$  of the decaying solution of the radial equation $X_m(\xi)$ \eref{small_X}.

\section{Spheroidal functions}\label{spheroidal_functions}

A convenient method to find periodic functions $Y_m(\gamma)$ is to expand them into a series of  Gegenbauer's polynomials \cite{meixner}
\begin{equation}
Y_m(\gamma)=\sum_{n\equiv m\, \mathrm{mod}\, 2} b_n C_n^{-\nu}(\cos \gamma)
\label{sum_Y}
\end{equation}
where the  summation is performed over all non-negative integers of the same parity as $m$.
 
Here $C_n^{\mu}(x)$ are polynomial solutions of the equation
\begin{equation}
(1-x^2)y^{\prime \prime}-(2\mu+1)xy^{\prime}+n(n+2\mu)y=0 
\end{equation}
and $C_n^{\mu}(\cos \gamma)$ are solutions of 
\begin{equation}
y^{\prime \prime}+2\mu \cot \gamma \, y^{\prime }+n(n+2\mu)y=0\ . 
\end{equation}
We impose the standard normalization \cite{bateman}
\begin{equation}
C_n^{\mu}(1)=\frac{\Gamma(n+2\mu)}{n!\Gamma(2\mu)}
\end{equation}
so the orthogonality relation for Gegenbauer's polynomials  is
\begin{equation}
\int_{-1}^{1}(1-x^2)^{\mu-1/2}C_n^{\mu}(x)C_m^{\mu}(x)\mathrm{d}x=\delta_{nm}\frac{2^{1-2\mu}\pi\Gamma(n+2\mu)}{n![\Gamma(\mu)]^2(n+\mu)}\ .
\label{norm}
\end{equation} 
Notice that \cite{bateman}
\begin{equation}
\mu C_n^{\mu}(\cos \gamma)\underset{\mu\to 0}{\longrightarrow} \frac{2}{n}\cos (n\gamma) 
\end{equation}
so for $\mu\to 0$ series \eref{sum_Y} can be transformed to series used for even Mathieu functions. 

Using the recurrence relation for Gegenbauer polynomials \cite{bateman} 
\begin{equation}
(n+1)C_{n+1}^{\mu}(x)=2(n+\mu)x C_{n}^{\mu}(x)-(n+2\mu-1)C_{n-1}^{\mu}(x)
\end{equation}
gives
\begin{equation}
\cos 2\gamma C_n^{-\nu}(\cos \gamma)=A_n \, C_n^{-\nu}(\cos \gamma)+B_{n-2}\, C_{n-2}^{-\nu}(\cos \gamma)+D_{n+2}\, C_{n+2}^{-\nu}(\cos \gamma)
\end{equation}
where 
\begin{equation}
A_n=-\frac{\nu(1+\nu)}{(n-\nu)^2-1}\ ,\;
B_n= \frac{(n-2\nu+1)(n-2\nu)}{2(n-\nu + 2)(n-\nu+1)}\ ,\;
D_n= \frac{(n-1)n}{2(n-\nu -2)(n-\nu-1)}\ . 
\end{equation}
Substituting the formal series \eref{sum_Y} into \eref{spheroidal} 
leads to a three-diagonals matrix  for the determination  of $b_n$ with fixed parity
\begin{equation}
\Big [\tfrac{1}{2}\theta^2 A_n -n(n-2\nu)\Big ] b_n+\tfrac{1}{2}\theta^2 B_n b_{n+2}+\tfrac{1}{2}\theta^2 D_n b_{n-2}=\alpha_m b_n\ . 
\end{equation}
Eigenvalues of the above matrix determine separation constants $\alpha_m$.

From \eref{norm} it follows that the normalization constant for this solution is
\begin{equation}
N_m=\sum_{n\equiv m \, \mathrm{mod}\,2}b_n^2 \frac{2^{1-2\mu}\pi\Gamma(n+2\mu)}{n![\Gamma(\mu)]^2(n+\mu)}\ .
\end{equation}
The function $\mathrm{e}^{ \theta x }$ is a solution of \eref{partial}. Therefore it can be expanded in elliptic coordinates \eref{elliptic_coordinates} as
\begin{equation}
\mathrm{e}^{ \theta   \cosh \xi \cos \gamma }=\sum_{m} Y_m(\gamma) \tilde{X}_{m}(\xi) 
\end{equation}
where $\tilde{X}_{m}(\xi)$ are  certain solutions (to be discussed  below) of the modified equation \eref{modified}.

Using \eref{orthogonality} one gets
\begin{equation}
\tilde{X}_{m}(\xi) =
\frac{1}{N_m}\int_0^{\pi}(\sin \gamma)^{-2\nu} \mathrm{e}^{ \theta \cosh \xi \cos \gamma} Y_m(\gamma)  \mathrm{d}\gamma \ .
\end{equation}
Due to the symmetry properties \eref{symmetry} 
\begin{equation}
\tilde{X}_{2m}(\xi) =
\frac{1}{N_{2m}}\int_0^{\pi}(\sin \gamma)^{-2\nu} \cosh [ \theta \cosh \xi \cos \gamma ]
 Y_{2m}(\gamma)  \mathrm{d}\gamma
\end{equation} 
and 
\begin{equation}
\tilde{X}_{2m+1}(\xi) =
\frac{1}{N_{2m+1}}\int_0^{\pi}(\sin \gamma)^{-2\nu} \sinh[ \theta \cosh \xi \cos \gamma]  Y_{2m+1}(\gamma)  \mathrm{d}\gamma\ .
\end{equation} 
Using  Gegenbauer's integral \cite{bateman}
\begin{equation}
n! \int_0^{\pi}\mathrm{e}^{\mathrm{i}z\cos \gamma}C_n^{\mu}(\cos \gamma)(\sin \gamma)^{2\mu}\mathrm{d}\gamma=
2^{\mu}\sqrt{\pi}\Gamma(\mu+1/2)\Gamma(n+2\mu)\mathrm{i}^{n}z^{-\mu}J_{n+\mu}(z)
\end{equation}
one obtains 
\begin{equation}
\tilde{X}_{m}(\xi)=\frac{2^{-\nu }\Gamma(1/2-\nu)}{N_{m}\Gamma(-2\nu)}\sum_{n\equiv m \, \mathrm{mod}\,   2} b_{n}\frac{\Gamma(n-2\nu)}{n!}(\theta \cosh(\xi))^{\nu}I_{n-\nu}(\theta \cosh(\xi))
\label{first_kind}
\end{equation}
where $I_{\mu}(x)$ are modified Bessel functions of the first kind.  

These functions increase exponentially at large $\xi$ and the series \eref{first_kind} represent  a solution of \eref{modified} of the first kind. As $I_{n-\nu}$ and $(-1)^nK_{n-\nu}$ obey the same recurrent relations, the exponentially decaying solutions  called solutions of the third kind  (denoted in \ref{series_solution} by $X_m(\xi)$) take the form
\begin{equation}
X_{m}(\xi)=\frac{2^{-\nu }\Gamma(1/2-\nu)}{N_{m}\Gamma(-2\nu)}\sum_{n\equiv m\, \mathrm{mod}\, 2} (-1)^n b_{n}\frac{\Gamma(n-2\nu)}{n!}(\theta \cosh(\xi))^{\nu}K_{n-\nu}(\theta \cosh(\xi))\ . 
\label{third_kind}
\end{equation}


\end{document}